\documentclass{cimento}

\newcommand{\kcpppn}{K^{\pm} \to \pi^0\pi^0\pi^{\pm}~}
\newcommand{\kcpp}{K^{\pm} \to \pi^{\pm}\pi^0~}
\newcommand{\kcppg}{K^{\pm} \to \pi^{\pm}\pi^0\gamma}
\newcommand{\ke}{K_{e 2}}
\newcommand{\km}{K_{\mu 2}}

\newcommand{\Kpe}{K^{\pm}\to{\pi^0 e^{\pm} \nu}~}

\newcommand{\RK} {\Gamma(K^{+}\to e^{+}\nu) /  \Gamma(K^{+}\to \mu^{+}\nu)}
\newcommand{\kcpnn}{K^{+} \to \pi^{+} \nu \bar{\nu}}

\usepackage{graphicx} 
\title{Rare K Decays: Present and perspectives with NA62}
\author{V.~Kozhuharov\from{ins:lnf}\from{ins:sofia} \thanks{Speaker, for
the NA62 Collaboration: F.~Ambrosino, A.~Antonelli, G.~Anzivino, R.~Arcidiacono, W.~Baldini, S.~Balev,
S.~Bifani, C.~Biino, A.~Bizzeti
% ...
 }
}

\instlist{
  \inst{ins:lnf} Laboratori Nazionali di Frascati - INFN, Frascati (Rome), Italy
  \inst{ins:sofia} University of Sofia ``St. Kl. Ohridski'', Sofia, Bulgaria
}

\PACSes{
  \PACSit{13.20.Eb}{Decays of K mesons}
  \PACSit{11.30.Hv}{Flavor symmetries}
  \PACSit{12.15.Hh}{Determination of Cabibbo-Kobayashi \& Maskawa (CKM) matrix elements}
}

\setcopyright{\copyright COPYRIGHT CERN on behalf of the NA62 Collaboration under CC BY-NC 3.0}

\begin{document}

\maketitle

\begin{abstract}
Rare kaon decays provide unique opportunity to test the Standard Model and probe its possible extensions. 
The final result on the lepton universality test by measuring the ratio $R_K = \RK$ is presented 
as well as the status of the study of the rare decays $K^\pm \to \pi^{\pm}\gamma\gamma $ and $K^{+} \to e^{+}\nu\gamma$. 
The primary goal of the NA62 experiment is the measurement of the  $Br(\kcpnn)$ decay with a precision of 10\% in two years of data taking. 
The detector setup together with the analysis technique is described.

\end{abstract}

\section{Introduction}

In the LHC era the flavour physics continues to play an important role in the 
determination of the properties of the fundamental particles. After the discovery of
the last particle of the Standard Model \cite{bib:higgs} one of the present goals of particle
physics is the search for a theory incorporating the knowledge from the cosmology - 
namely the matter-antimatter 
asymmetry and the invisible content of the Universe. Another focus is also given to the 
QCD and especially to its non-perturbative regime. 

The kaon physics with its precision allows to probe both the low energy behaviour 
of the strong interactions as well as the high energy weak scale through loop processes. 
Special attention should be given to the rare kaon decays since some of them 
proceed as flavour changing neutral currents while others have low rates due to helicity suppression. 
Moreover they could achieve sizeable contribution in the presence of New Physics.

\section{NA62 experiment}
The continuation of the CERN long standing kaon physics program is the NA62 experiment.
Its first phase took place in the 2007-2008 with the NA48/2 setup and was devoted 
to the study ot the $\ke$ decays.  In 2009 the existing
experimental apparatus was dismantled in order to allow the construction of the new setup \cite{bib:na62tdr} devoted to the 
study of the $\kcpnn$ decay. In 2012 two different test runs were accomplished: a dry run  in the summer devoted to the setup 
of the readout electronics, trigger and data acquisition chain and a technical run in the autumn with beam to study 
the performance of part of the NA62 subdetectors.

  \subsection{Phase I}
 The kaon beam was formed by a primary 400 GeV/c proton beam extracted from SPS hitting a 
400 mm long beryllium target. 
The secondary particles were selected with a momentum of $(74 \pm 1.4)$ GeV/c with the possibility 
to use simultaneous or single positive and negative beams. The fraction of kaons in the beam 
was about 6\% and they decayed in a 114 m long evacuated tank. 

The decay products were registered by the NA48 detector \cite{bib:na48}. The momentum 
of the charged particles was measured with resolution $\sigma(p)/p = (0.48 \oplus 0.009p[GeV/c]) )\%  $ 
by a spectrometer consisting of four drift chambers separated by a dipole magnet. Precise time information and trigger
condition was provided by a scintillator hodoscope with time resolution of 150 ps which was followed by a quasi-homogeneous liquid
krypton electromagnetic calorimeter measuring photon and electron energy with 
resolution $\sigma(E)/E = 3.2\%/ \sqrt{E} \oplus 9\%/ E \oplus  0.42\%$ [GeV]. It was also able to 
provide particle identification based on the energy deposit by different particles with respect to 
their momentum.
 
  \subsection{Phase II}
  The beam and the detector for the second phase of the NA62 experiment are new and 
 are dictated by the main goal - the study of the extremely rare decay $\kcpnn$. The protons intensity from the SPS will 
be increased by 30\% and the secondary positive beam will be with momentum $(75GeV/c  \pm 1\% )$. Its rate will be 
about 800 MHz and the decay volume is evacuated. The final beam line was tested during the technical run in 2012. 
\begin{figure}
 \includegraphics[height=.18\textheight]{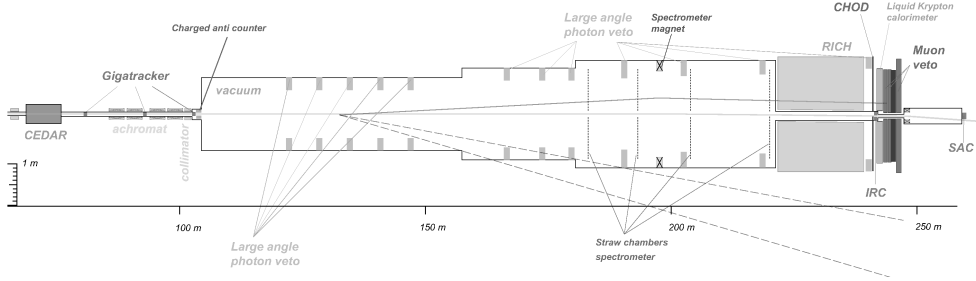}
 \caption{NA62 experimental layout}
 \label{fig:na62}
\end{figure}
  The major detector components are shown in fig. \ref{fig:na62} and are:

{\bf CEDAR (KTAG):} Hydrogen filled threshold Cherenkov counter used for positive kaon identification in the beam.
Time resolution of 100 ps was achieved during the technical run which is important for the suppression of accidental
background. 

{\bf Gigatracker:} Three stations of thin silicon pixel detectors for the measurement of kaon momentum,
 flight direction and time.
The expected resolutions on the measured quantities will be $\sigma(p_K)/p_K \sim 0.2\%$ on momentum, 16 $\mu rad$ angular and  time resolution of 200 ps per station. 

{\bf Chanti:} Scintillating anticounters providing a veto against interactions of the beam particles. 

{\bf ANTI:} Twelve rings of lead glass counters surrounding the decay region and acting as photon veto detectors (LAV) for angles
of the photons higher than 15 $mrad$ with respect to the kaon flight direction. Currently eight LAVs were built and two of them were successfully
operated in 2012. 

{\bf Straw spectrometer:} Four chambers of straw tubes separated by the MNP33 dipole magnet will be operated in vacuum in order to provide
momentum resolution of $\sigma(p)/p = (0.3 \oplus 0.007 p[GeV/c]) )\%  $ with a minimal material budget. Half a chamber was tested
during the technical run.

{\bf RICH:} Ring imaging Cherenkov detector will measure the velocity of the charged particles allowing to separate 
pions from muons and will provide time resolution better than 100 ps \cite{bib:rich}. 

{\bf CHOD:} The charged hodoscope will be fast plastic scintillator counter detector and will be used in the trigger. 

{\bf IRC and SAC:} Shashlyk type veto detector covering photon angles down to zero (SAC) and also serving 
to veto converted in the upstream material photons (IRC). SAC was operated successfully in 2012 with a stand alone 
gigahertz flash ADC readout. 

{\bf LKr:} The NA48 liquid krypton calorimeter with renewed readout electronics will serve as a 
photon veto for photons with angles from 1 to 15 $mrad$ with inefficiency less than 
$10^{-5}$ for photon with energies above 10 GeV. 

{\bf MUV:} Three muon veto stations based on iron and scintillator sandwich 
will provide separation between pions and muons better than $10^{-5}$. 

Both KTAG and Gigatracker are exposed to the full 800 MHz hadron beam while the rate by the downstream detectors is 
at most 10 MHz.

\section{Results on lepton universality}

The dilepton charged pseudoscalar meson decays proceed as tree level 
processes within the Standard Model through a W exchange. 
However the electron mode is strongly suppressed by the helicity conservation. 
The theoretical expression for the ratio $R_K= \Gamma(Ke2) / \Gamma(K\mu 2)$ is given by
\begin{equation}
R_K=\frac{m_e^2}{m_{\mu}^2} \left(  \frac{m_K^2 - m_e^2}{m_K^2 - m_{\mu}^2} \right)   (1+\delta R_K)
\end{equation}
where the term $ \delta R_K = -(3.79 \pm 0.04) \% $ represents the radiative corrections. 
In the ratio $R_K$ the theoretical uncertainties on the hadronic matrix element cancel 
resulting in a precise prediction $R_K = (2.477 \pm 0.001) \times 10^{-5} $ \cite{ke2-thnew}. 

In various extensions of the SM (like models with Lepton Flavour Violation, 
different two Higgs doublet models) a constructive or destructive contribution to $R_K$ as high as \% could be achieved \cite{Masiero}. 

Within the NA48/2 experiment two preliminary measurements were performed \cite{ke2-na48}  
which allowed to optimize the analysis strategy. The final NA62 result, using data from phase one of the experiment,
 supersedes the NA48/2 ones and 
the preliminary NA62 result which was based on the analysis of a partial 
data sample collected in 2007 \cite{bib:na62-ke2prel}. 

From experimental point of view the ratio $R_K$ can be expressed as
\begin{equation}
R_K =
\frac{1}{D}\cdot
\frac{N(\ke)-N_B(\ke)}
{N(\km) - N_B(\km)}
\cdot
\frac{A(\km)\times\epsilon_{\mathrm{trig}}(\km)\times f_\mu}
{A(\ke)\times\epsilon_{\mathrm{trig}}(\ke)\times f_e} \cdot \frac{1}{f_{LKr}},
\label{RKexp}
\end{equation}
where  $N(K_{\ell 2})$, $\ell=e,\mu$ is the number of the selected $\ke$ 
and $\km$ candidates,  
$N_B(K_{\ell 2})$ is the number of expected background events, 
$f_\ell$ is the efficiency for particle identification, 
$A(K_{\ell 2})$ is the geometrical efficiency for registration obtained from Monte Carlo simulation, 
$\epsilon_{\mathrm{trig}}$ is the trigger efficiency, $D=150$ is the downscaling factor for $\km$ events and $f_{lkr}$ is the global efficiency of the LKR readout. Both $f_\ell$ and $\epsilon_{\mathrm{trig}}$ are higher than 99\%. 
The analysis is performed in individual momentum bins for all the four different data taking conditions resulting into 
40 independent values for the $R_K$. 

The similarity between the two decays allowed to exploit systematics cancellations by using common selection criteria. The events were required to have only one reconstructed charged track consistent with kaon decay within the detector geometrical acceptance with momentum in the interval $13 ~GeV/c < p < 65 ~GeV/c$. 
In order to suppress the background a cut on extra non-associated with the track LKr clusters with energy above 2 $GeV$ was made. 
The particle identification was based on the $E/pc$ variable, where $E$ is the energy deposited in the LKr and $p$ is the momentum measured by the spectrometer. 
It had to be close to one for electrons and less than 0.85 for muons. 
Under the assumption of the particle type the missing mass squared was calculated $M_{miss}^2 = (P_K - P_l)^2$, 
where $P_K$ ($P_l$) is the kaon (lepton) four momentum. A momentum dependent cut on the $M_{miss}^2$  was used.

The dominant background contribution in the $\ke$ sample was identified to come from $\km$ events with muons leaving all their energy in the electromagnetic calorimeter. 
The two decays are well separated below 35 $GeV/c$ track momentum but completely overlap kinematically for higher values. 
In order to select clean muons a sample of data was taken with a $9.2 X_0$ thick Pb wall placed in front of the LKr. It was used to estimate the amount of muons faking electrons. At low track momentum the most significant background source was identified to be the muon halo. The total background was $ (10.95 \pm 0.27)\%$. 

\begin{figure}[!htb]
    \resizebox{0.42\textwidth}{!}{\includegraphics[width=0.42\textwidth]{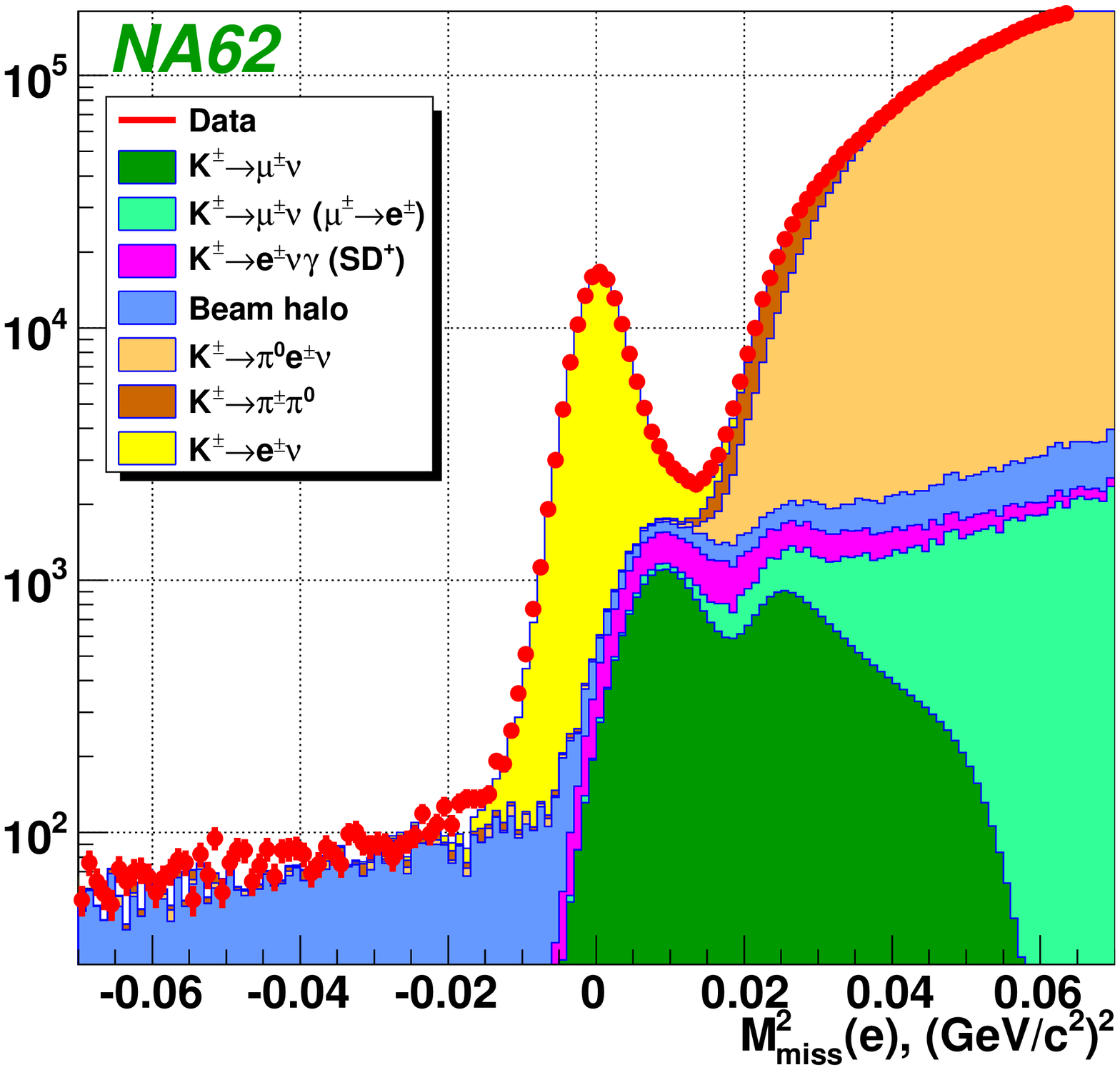}}
    \resizebox{0.49\textwidth}{!}{\includegraphics[width=0.49\textwidth]{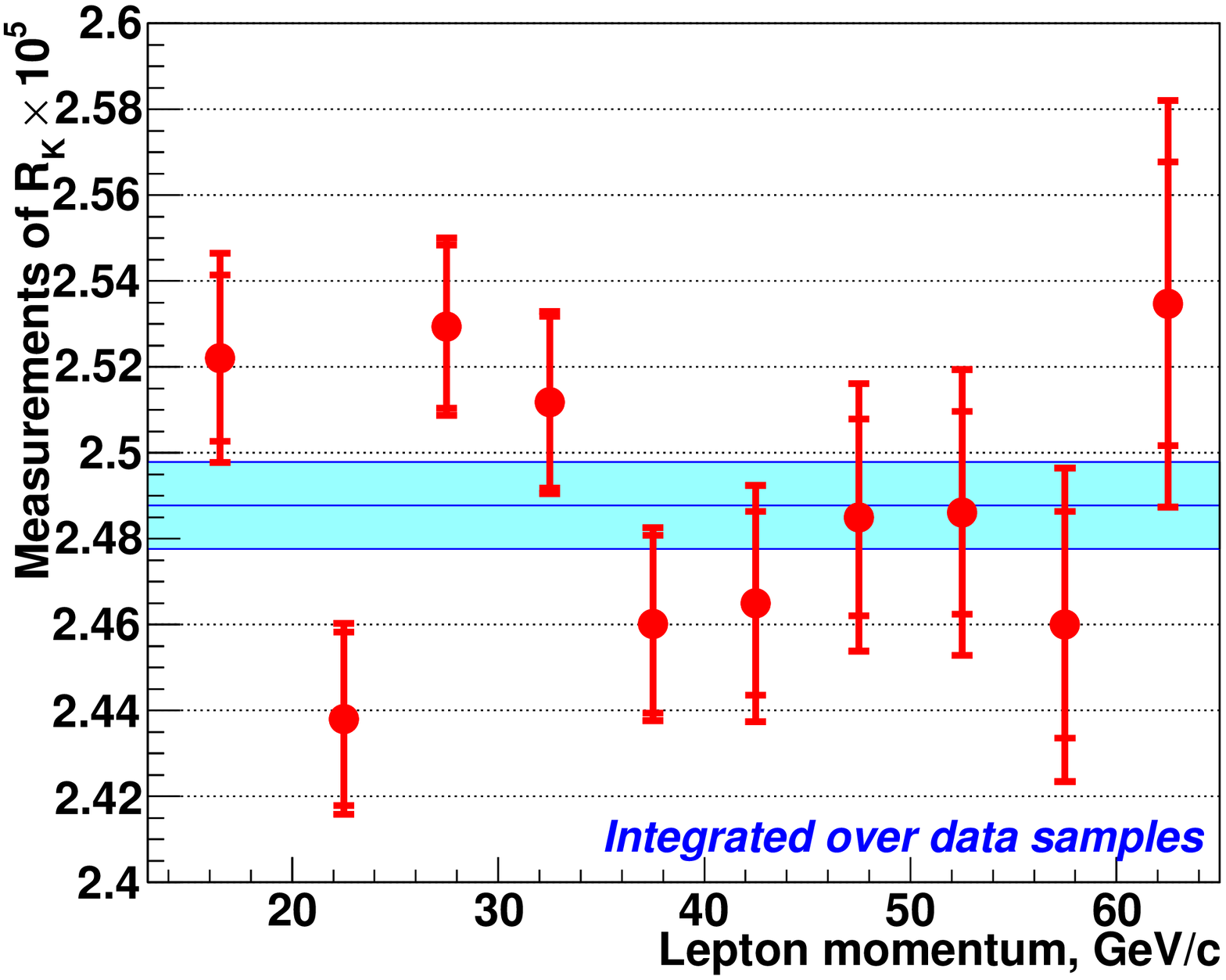}}
\put(-350,161){ \bf (a)}
\put(-150,161){ \bf (b)}
    \caption{ {(a) Squared missing mass distribution for the reconstructed data events (dots) together with the major background contribution. (b) $R_K$ in bins of track momentum integrated over data samples. The horizontal dashes represent the statistical only and the total error.} \label{fig:ke2} }
\end{figure}

The missing mass distribution for the reconstructed $\ke$ data events together with the simulation of the signal and backgrounds are shown in fig. \ref{fig:ke2}(a). A total of $145958$ $\ke$ and $4.28\times 10^{7}$  $\km$ candidates were reconstructed. The value of $R_K$ in the in individual momentum bins integrated over the data samples is shown in fig. \ref{fig:ke2}(b). The final result was obtained by a fit of the 40 independent $R_K$ values and is
\begin{equation}
  R_K = (2.488 \pm 0.007_{stat} \pm 0.007_{syst})\times 10^{-5}.
\end{equation}
It is consistent with the Standard Model prediction and with the present PDG value \cite{bib:ke2-pdg}.

\section{Studying chiral perturbation theory}
  Radiative kaon decays provide a crucial test of the ability of the Chiral Perturbation Theory (ChPT) to describe low energy 
weak processes.

  \subsection{$K^{\pm}\to \pi^{\pm}\gamma\gamma$ decay}
  
  Within the ChPT the lowest order terms contributing to the the decay $K^{\pm}\to \pi^{\pm}\gamma\gamma$ are of order $O(p^4)$ \cite{bib:pigg-th1}. 
They represent the pion and the kaon loop amplitudes depending on a single unknown constant $\hat{c}$ and a pole amplitude contributing of the 
order of 5\% to the final decay rate. 
Higher order corrections ($O(p^6)$) have shown to change the decay spectrum significantly leading to a non vanishing differential decay rate at zero 
diphoton invariant mass \cite{bib:pigg-th2}. The predicted branching fraction for the $K^{\pm}\to \pi^{\pm}\gamma\gamma$ is $\sim 10^{-6}$ and has been studied by the BNL E787 experiment which observed 31 decay candidates \cite{bib:pigg-bnl}. 

The decay $K^{\pm}\to \pi^{\pm}\gamma\gamma$ was studied using a minimum bias sample collected in 
three day NA48/2 run in 2004 and during the 2007 NA62 data taking. The kaon momentum in the NA48/2 data was 
$ (60 \pm 3)$ GeV/c while the effective trigger downscaling in the 2007 run was about 20. This resulted in 
similar kaon fluxes but different resolution and background conditions. 

\begin{figure}[!htb]
    \resizebox{0.49\textwidth}{!}{\includegraphics[width=0.49\textwidth]{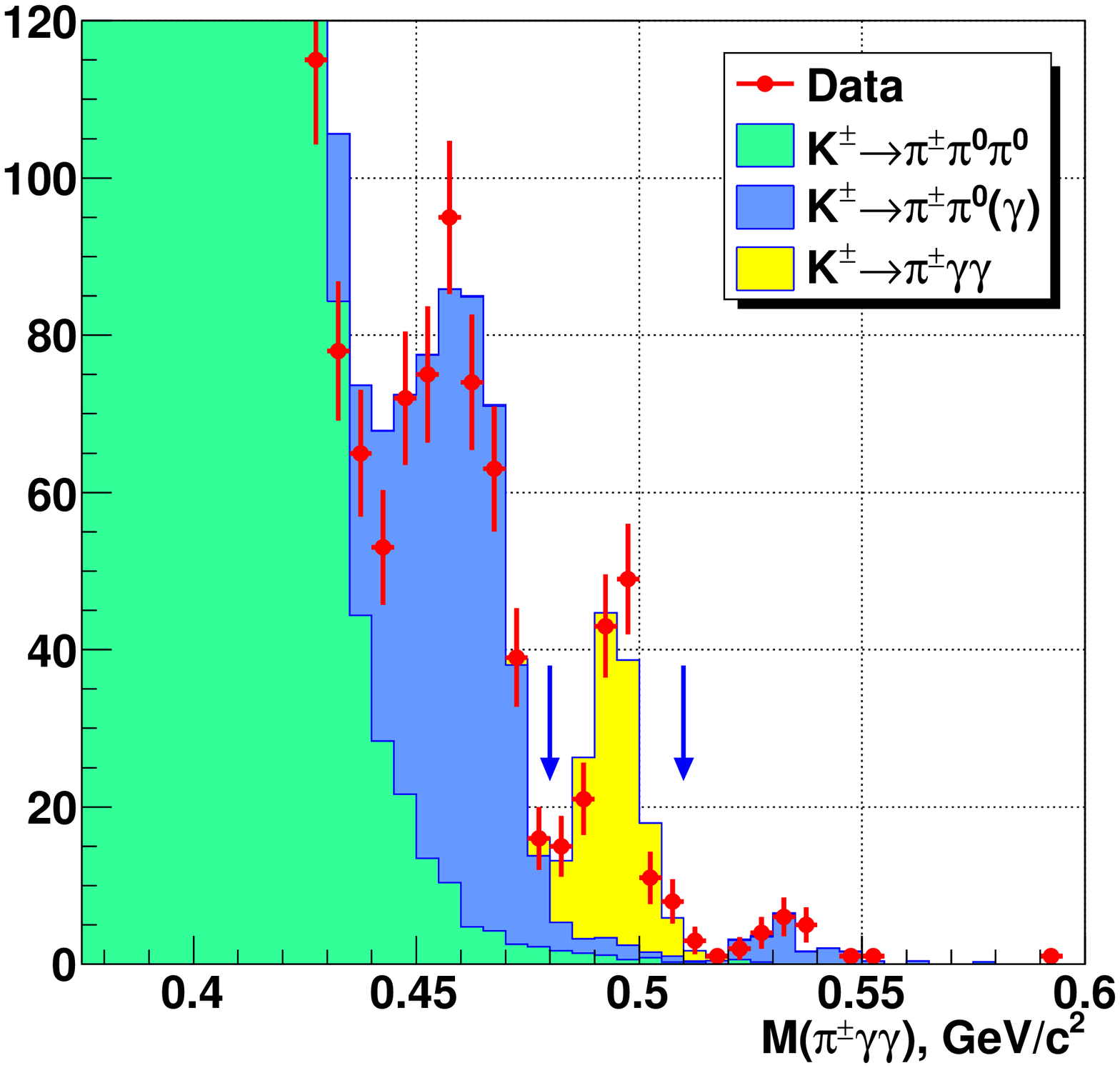}}
    \resizebox{0.49\textwidth}{!}{\includegraphics[width=0.49\textwidth]{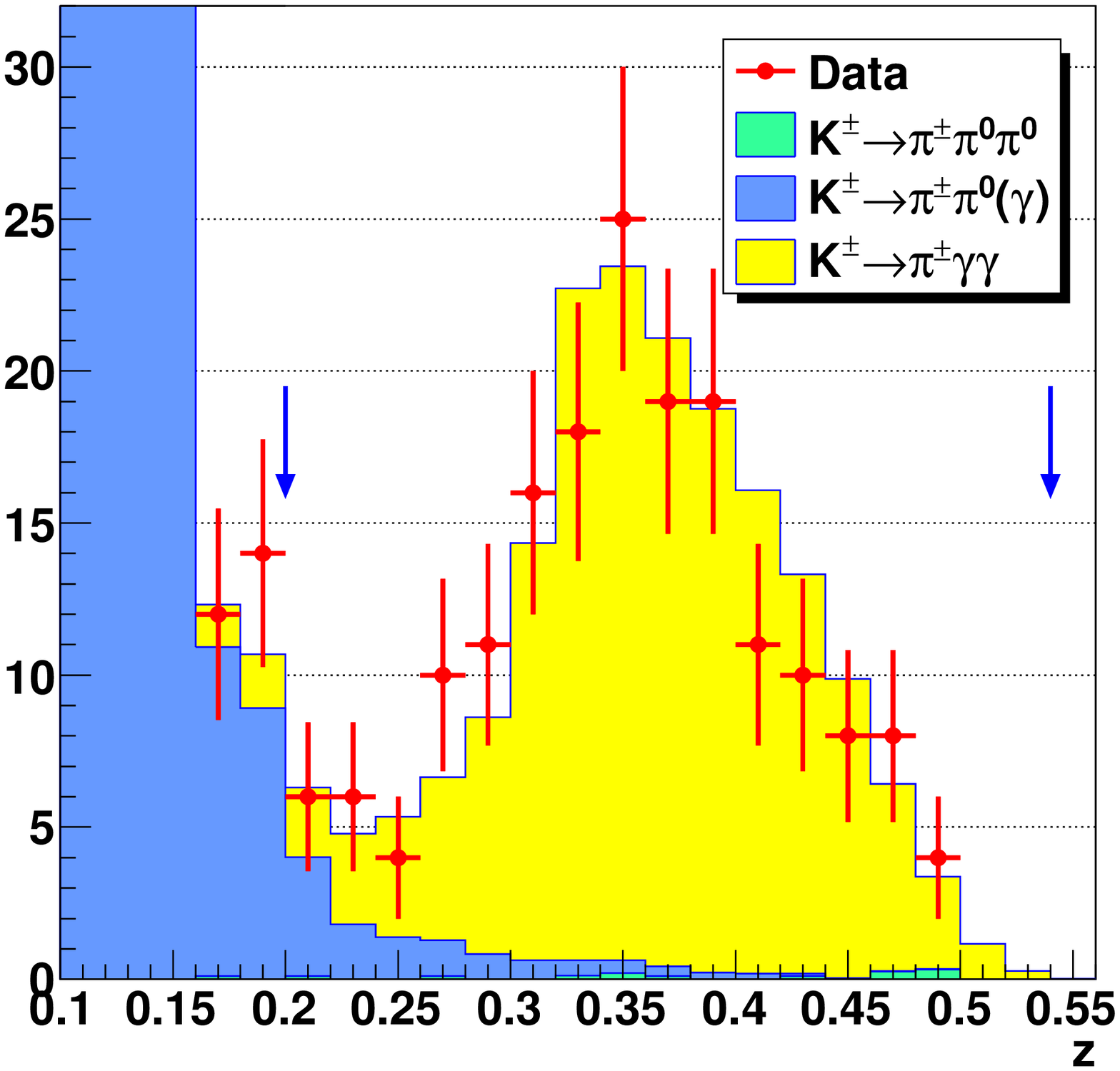}}
\put(-350,180){ \bf (a)}
\put(-150,180){ \bf (b)}
    \caption{ {(a) Invariant mass for the reconstructed data events in the 2004 data together with the expected background contribution. (b) The data spectra of the z kinematic variable for the 2007 data sample with the expected background 
}  \label{fig:kpgg} }
\end{figure}

The signal events were required to have $z= (m_{\gamma\gamma} / m_K)^2 > 0.2$ in order to suppress the background from
$\kcpp$ decays. The reconstructed kaon invariant mass for the 2004 event candidates is shown in fig. \ref{fig:kpgg} (a). 
A total of 147 (175) event candidates were selected in the 2004 (2007) data set, with backgrounds
contaminations of 12\% (7\%) from $\kcpppn$ and $\kcppg$ with merged clusters. The $z$ spectrum for the 2007 event
candidates is shown in fig. \ref{fig:kpgg} (b). The extraction of the $\hat{c}$ was based on a likelihood fit to the data
for both $O(p^4)$ and $O(p^6)$ parametrizations. The preliminary results for $\hat{c}$ together with the branching fraction for the $K^{\pm}\to \pi^{\pm}\gamma\gamma$ decay 
 assuming $O(p^6)$ parametrizations are shown in table \ref{tab:pigg-prel}.

\begin{table}[!htb]
  \caption{Preliminary results for $\hat{c}$ and $BR(K^{\pm}\to \pi^{\pm}\gamma\gamma)$ within ChPT parametrization. }
  \label{tab:pigg-prel}
  \begin{tabular}{c|c|c|c}
    \hline
           & $\hat{c}$, $O(p^4)$                      & $\hat{c}$, $O(p^6)$                    & $BR(K^{\pm}\to \pi^{\pm}\gamma\gamma)$, $O(p^6)$\\
    \hline
2004 data  &  $1.36 \pm 0.33_{stat} \pm 0.07_{syst}$  & $1.67 \pm 0.39_{stat} \pm 0.09_{syst}$ &  $ (0.94 \pm 0.08) \times 10^{-6}$  \\
2007 data  &  $1.71 \pm 0.29_{stat} \pm 0.06_{syst}$  & $2.21 \pm 0.31_{stat} \pm 0.08_{syst}$ & $(1.06 \pm 0.07) \times 10^{-6}$ \\
    \hline
Combined   &  $1.56 \pm 0.22_{stat} \pm 0.07_{syst}$  & $2.00 \pm 0.24_{stat} \pm 0.09_{syst}$ & $(1.01 \pm 0.06) \times 10^{-6}$ \\
    \hline
  \end{tabular}
\end{table}

Statistical error is dominant while the major contribution to the systematics comes from the background estimation. 
The total uncertainty on the result doesn't allow to discriminate between the two parametrization. 
Moreover at $\hat{c} \sim 1.5$ both $O(p^4)$ and $O(p^6)$ give the same value for the $BR(K^{\pm}\to \pi^{\pm}\gamma\gamma)$.

  \subsection{$K^+\to e^+\nu\gamma$ decay}

\begin{figure}[!htb]
    \resizebox{0.43\textwidth}{!}{\includegraphics[width=0.43\textwidth]{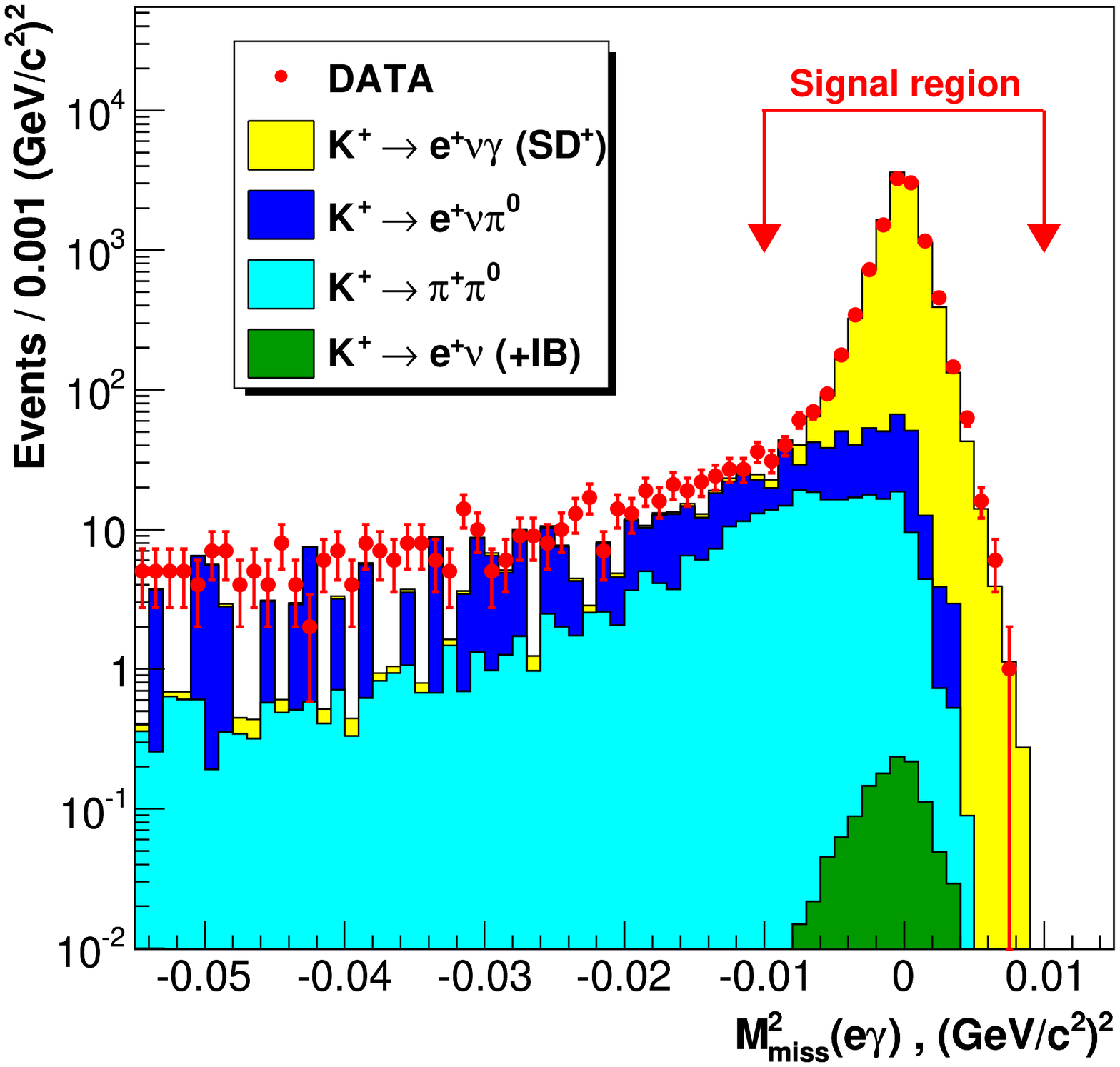}}
    \resizebox{0.55\textwidth}{!}{\includegraphics[width=0.55\textwidth]{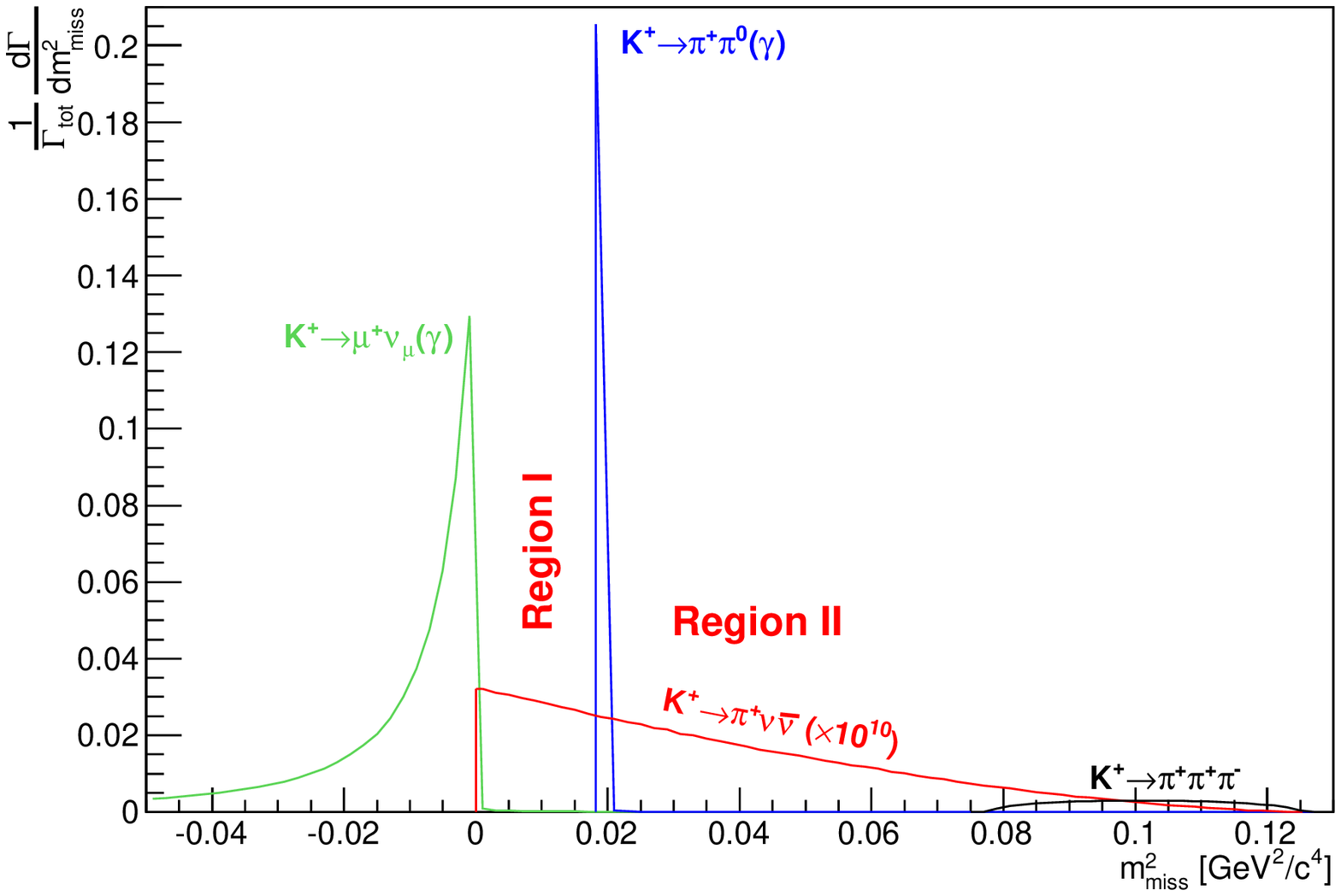}}
\put(-350,161){ \bf (a)}
\put(-150,161){ \bf (b)}

    \caption{ {(a) Squared missing mass distribution for the $K^+\to e^+\nu\gamma$ candidates. (b) Squared missing mass distribution for the $\kcpnn$ and the kinematically limited kaon decays.}     \label{fig:kna62} }
\end{figure}

The contributions to the $K^+\to e^+\nu\gamma$ decay amplitude can be divided into inner bremsstrahlung (IB) 
and structure dependent (SD) part. The IB is purely electromagnetic and is included in the definition of the 
$R_K$ while the SD term is sensitive to the kaon structure. It can be parametrized with two form-factors - vector $F_V$ 
and axial vector $F_A$. The NA62 experiment collected about 10000 event candidates in its first phase of data taking which 
represents about an order of magnitude higher statistics than the recently published result by KLOE collaboration 
\cite{bib:ke2g-kloe}. 
The major background contributions were identified to originate from $\Kpe$ with $\pi^0 \to e^+e^-\gamma$ and $\kcpp$ decays. 
The reconstructed neutrino missing mass for the data events and the simulated signal and background are shown in fig. \ref{fig:kna62}(a). 

The analysis is ongoing and a ChPT $O(p^6)$ fit is being performed on the data. NA62 collaborations aims also at a model independent
extraction of the form factors.

\section{Measurement of the $Br(K^+ \to \pi^+\nu\bar{\nu})$ with 10\% precision}
Among the rare kaon decays the transitions $K \rightarrow \pi \nu \bar{\nu}$ are extremely attractive. 
They proceed as FCNC and their branching fractions are theoretically very clean since the hadronic 
matrix element can be obtained by the isospin symmetry of the strong interactions from the leading 
decay $K \rightarrow \pi e \nu$ \cite{ISOSPIN_RELATION}. For the charged kaon mode the NNLO 
calculations give $Br(\kcpnn) = (7.81\pm0.80)*10^{-11}$ \cite{pnn-th}. 

Presently seven $\kcpnn$ events have been observed by the E787 and E949 collaborations 
in a stopped kaon experiment \cite{BNL_BR} leading to $Br(\kcpnn) = (1.73_{-1.05}^{+1.15})\times 10^{-10}$. 
This value is twice the SM prediction but still compatible with it due to high uncertainty. 
The decay $\kcpnn$ is very sensitive to New Physics models where the theoretical predictions vary over an order of magnitude. 
Thus measuring $Br(K^+ \rightarrow \pi^+ \nu \bar{\nu})$ could help to distinguish between the different 
types of New Physics \cite{PINN_BSM} once it is discovered.

The measurement NA62 collaboration aims to perform is based on the kaon decay in flight. 
In order to identify the signal events and suppress the background the three techniques - 
kinematics, particle identification and vetoing will be exploited.
Since in the final state there is only one observable particle the kinematics
variable considered for the separation of the decay is the missing mass squared under 
pion hypothesis for the charged track. It is defined as 
\begin{equation}
 m_{miss}^2 \simeq m_K^2\left(1-\frac{|P_{\pi}^2|}{|P_{K}^2|}\right)  + m_{\pi}^2\left(1-\frac{|P_{K}^2|}{|P_{\pi}^2|}\right) - |P_{K}||P_{\pi}|\theta_{\pi K}^2 
\end{equation}
With the planned Gigatracker and Straw spectrometer the expected resolution on the missing mass squared is $0.001 ~GeV^2/c^4$. 
The signal region is defined by the edges of the kinematically limitted in the $m_{miss}^2$ 
distribution kaon decays: $K^+ \to \mu^+ \nu$, $K^+ \to \pi^+ \pi^0$ and $K^+\to \pi^+\pi^+\pi^-$, as shown in fig. \ref{fig:kna62}(b). 
The detector is almost hermetic for photons originating from $\pi^0$ in the decay region and 
overall photon veto system composed from ANTI, LKR, SAC, IRC provides an inefficiency less than $10^{-8}$ for a $\pi^0$ coming 
from $K^+\rightarrow \pi^+\pi^0$ decay. The conversions of low energy photons in the upstream material (RICH, beam tube) 
will be detected by the CHOD.

Decays with muons in the final state (like $K^+\rightarrow \mu^+ \nu$, $K^+\rightarrow \pi^+\pi^-\mu^+ \nu$) 
will be suppressed using the muon-pion identification based on RICH and MUV for which the total 
inefficiency should be less than  $5 \times 10^{-6}$.

The presented setup and analysis strategy will allow NA62 experiment to collect O(100) events in two years of data taking. 
The construction is within the timeline for the start of the experiment in 2014.

\section{Conclusion}
The rare kaon decays continue to provide a valuable input to the high energy physics. 
The NA62 experiment with its huge statistics, excellent resolution, particle identification, and 
hermeticity will be the future laboratory for charged kaon physics. 
Currently a measurement of $R_K$ with four per mile uncertainty has been performed and a new higher statistics studies
of the $K^{\pm}\to \pi^{\pm}\gamma\gamma$ and $K^+\to e^+\nu\gamma$ decays are ongoing. 
The next step is the study of $\kcpnn$ decay and the measurement of the CKM matrix element
$V_{td}$ with a 10\% precision. 
The general purpose experimental setup could allow to study  
other rare decays, namely lepton flavour and lepton number violating ones. 
The simultaneous kaon and pion beams could also provide a unique opportunity 
to perform the first direct measurement of the ratio $R^e_{K\pi} = \Gamma(Ke2)/\Gamma(\pi e2)$ with 
0.3\% precision if it is possible to control the beam composition precisely.



%


\begin{thebibliography}{0}

\bibitem{bib:higgs}
  \BY{ G.~Aad {\it et al.}  [ATLAS Collaboration]} 
  \IN{Phys.\ Lett.\ B} {716} {2012} {1};
  \BY{ S.~Chatrchyan {\it et al.}  [CMS Collaboration] }
  \IN{Phys.\ Lett.\ B} {716} {2012} {30}.

\bibitem{bib:na62tdr} 
  NA62 Technical Design Document, NA62-10-07.

\bibitem{bib:na48} \BY{V. Fanti {\it et al.} [NA48 Collaboration]}
    \IN{Nucl. Instrum. Meth. A} {574} {2007} {433}.


\bibitem{bib:rich} \BY{B. Angelucci {\it et al.}}
  \IN{Nucl.\ Instrum.\ Meth.\ A} {621} {2010} {205}.
  

\bibitem{ke2-thnew} \BY{V. Cirigliano \atque I. Rosell}
  \IN{Phys. Rev. Lett.} {99} {2007} {231801}


\bibitem{Masiero} \BY{A. Masiero {\it et al.} }
\IN{Phys. Rev. D} {74} {2006} {011701}.

\bibitem{ke2-na48} \BY{L. Fiorini} 
\IN{PoS} {HEP2005} {2006} {288}; 
\BY{V.~Kozhuharov} \IN{PoS} {KAON} {2008} {049}.

\bibitem{bib:na62-ke2prel} \BY{C.~Lazzeroni {\it et al.}  [NA62 Collaboration]}
\IN{Phys.\ Lett.\ B} {698} {2011} {105}

\bibitem{bib:ke2-pdg} \BY{J. Beringer  {\it et al.} [Particle Data Group]}
\IN{Phys. Rev. D} {86} {2012} {010001}.


\bibitem{bib:pigg-th1} \BY{G. Ecker, A. Pich \atque E. de Rafael}
\IN{Nucl. Phys. B} {303} {1988} {665}.


\bibitem{bib:pigg-th2} \BY{G. D’Ambrosio \atque J. Portoles}
\IN{Phys. Lett. B} {386} {1996} {403}.

\bibitem{bib:pigg-bnl} \BY{P. Kitching {\it et al.} [E787 Collaboration]}
\IN{Phys. Rev. Lett.} {79} {1997} {4079}.


\bibitem{bib:ke2g-kloe} \BY{F. Ambrosino {\it et al.} [KLOE Collaboration]}
\IN{Eur. Phys. J. C} {64} {2009} {627}


\bibitem{ISOSPIN_RELATION}
  \BY{F. Mescia \atque C. Smith}
\IN{Phys. Rev. D} {76} {2007} {034017}.

\bibitem{BNL_BR}
\BY{A. Artamonov  {\it et al.} [E949 Collaboration]}
\IN{Phys. Rev. D} {79} {2009} {092004}.

\bibitem{pnn-th}
\BY{J. Brod, M. Gorbahn \atque E. Stamou}
\IN{Phys. Rev. D} {} {2011} {034030}.

\bibitem{PINN_BSM}
\BY{D. Straub}
arXiv:1012.3893 [hep-ph]


\end{thebibliography}
\end{document}